\documentclass[11pt]{article}
\usepackage{moriond,epsfig}

\def\circa#1{\,\raise.3ex\hbox{$#1$\kern-.75em\lower1ex\hbox{$\sim$}}\,}
\newcommand{\ov}{{\cal O}}

 \newcommand \bra {\langle}
\newcommand \ket {\rangle} \newcommand{\be}{\begin{equation}}
\newcommand{\ee}{\end{equation}} \newcommand{\ben}{\begin{displaymath}}
\newcommand{\een}{\end{displaymath}} \newcommand{\ba}{\begin{eqnarray}}
\newcommand{\ea}{\end{eqnarray}} \newcommand{\ban}{\begin{eqnarray*}}
\newcommand{\ean}{\end{eqnarray*}} 
 \newcommand{\de}{\partial}

\newcommand{\kt}{\mbox{$k$}_{\perp }}

\newcommand{\effe}[1]
{\mathop{{\vspace{3pt}\cal F}}\limits^{}_{\hskip-1mm \scriptscriptstyle #1}}

\newcommand{\ff}[2]{\mathop{{\vspace{3pt}f}}\limits^{}_{\hskip-1mm
  \scriptscriptstyle #1}{\hskip-1mm \scriptstyle #2}}

\newcommand{\F}[2]
{\mathop{{\vspace{3pt}\cal F}}\limits^{#1}_{\hskip-1mm \scriptscriptstyle #2}}
\newcommand{\der}{-\frac{\de}{\de t}}
\newcommand{\overlap}[1]
{\mathop{{\vspace{3pt}\cal O}}\limits^{}_{\scriptscriptstyle #1}}

\begin{document}
\vspace*{4cm}
\title{EVOLUTION EQUATIONS IN THE ELECTROWEAK SECTOR OF THE STANDARD MODEL}

\author{PAOLO CIAFALONI }

\address{INFN  Sezione di Lecce, Via per Arnesano\\
73100 Lecce, Italia}

\maketitle\abstracts{Energy-growing
 electroweak corrections in the Standard Model are
potentially  relevant for LHC physics,
for Next generation of Linear Colliders (NLCs)
 and for ultrahigh energy cosmic
rays. I present here the results of recent work in which
 electroweak evolution equations (the analogous of DGLAP equations in QCD)
 have been derived.  The main features of these effects, mainly related to
 the fact that the electroweak sector is spontaneously broken,
 are pointed out.}

\section{Introduction}
The study of radiative electroweak corrections in the Standard Model 
at very high (TeV scale)
energies has taught us many interesting, mostly unexpected,
features.
 Such corrections turn out to be  huge, of the order of 10-20
\% at one loop, 
and grow like the square of the log of the c.m. energy. This has
triggered a fairly big amount of work on higher order corrections and on 
the possibility of resumming them \cite{LL}, 
particularly after the observation that
such logarithms are tied to the infrared structure of the theory \cite{cc}. 
While resummation techniques can be
mutuated from QCD, it has been clear from the beginning that the fact that
the electroweak SU(2) $\otimes$ U(1) sector is spontaneously 
broken causes important
differences from an unbroken theory like the SU(3) strong interactions sector.

The most striking  feature of electroweak radiative corrections at
asymptotic energies is
the following. Consider a very high energy process ($\sqrt{s}\gg 100$ GeV)
in which you include over all possible form of radiation in the final
state: photons, gluons, but also Ws and Zs. Suppose that every kinematical
variable defining the process is of the same order. A paradigm for this
kind of observable is the process $e^+e^-\to 2 jets+X$ with
$s\approx |t|\approx |u|$. Then, one would naively conclude that the cross
section for this process depends on the only kinematical variable $s$ and
that its asymptotic 
behaviour for large $s$ is dictated by the renormalization group
equations. However, this turns out to be false: even at the highest
energies, fully inclusive observables depend on the weak scale $M\approx
100$ GeV, which acts as an infrared cutoff for potentially divergent double
logarithms. This dependence is due to a lack of cancellation between real
and virtual emission, related to the fact that initial states are
nonabelian (isospin) charges  \cite{3p1}.

In this note I present the results of two papers in which the infrared
evolution equations in the Standard Model of electroweak interactions have
been derived. These equations correspond to the usual
Dokshitzer-Gribov-Lipatov-Altarelli-Parisi (DGLAP) equations in QCD,
however because of symmetry breaking in the electroweak sector a number of
differences emerge.
\begin{itemize}
\item
because of the above mentioned lack of cancellation, electroweak evolution
equations take into account not only single logarithmic collinear
contributions $\sim\alpha_W\log\frac{s}{M^2}$, but also double logarithms
of infrared/collinear origin  $\sim\alpha_W\log^2\frac{s}{M^2}$.
\item
new splitting functions, that are different from the ones appearing in QCD,
have to be introduced.
\item
while kinematics closely resembles that of QCD, the lack of isospin
averaging (compared with colour averaging in QCD) render the isospin
structure much more complicated.
\end{itemize}

Energy-growing electroweak corrections in the Standard Model are
potentially  relevant for LHC physics \cite{acco},
for Next generation of Linear Colliders (NLCs) \cite{nlc}
 and for ultrahigh energy cosmic
rays \cite{berez}.

\section{Left fermions in the $g'\to 0$ limit}

In this section I consider lepton initiated Drell Yan process
of type $e^+(p_1)\;e(p_2)\rightarrow q(k_1) \bar{q}(k_2)  +X$\footnote{note
that the process considered, like all others in this paper, is fully
inclusive, meaning also $W,Z$ radiation is included.} 
where $s=2 p_1 \cdot p_2$ is the total invariant mass and $Q^2=2 k_1 \cdot k_2$
 is the hard scale. I consider double log corrections in relation to the
 SU(2) electroweak gauge group, i.e. I work in the limit where the U(1)
 coupling   $g'$ is zero. This process 
has been analyzed in \cite{strong,full}, 
to which I refer for details; I consider it here as a simple example of how
evolution equations can be derived.
The general formalism used to study electroweak evolution equations for
inclusive observables has been set up in \cite{col}; I summarize it here
briefly. To begin with, by arguments of unitarity, final state
radiation can be neglected when considering inclusive cross sections
\cite{strong}. Then one is led to consider the dressing
of the overlap matrix $\overlap{}\!_{\alpha\beta}^{\alpha'\beta'}= 
\bra \beta\beta'|
S^+\;S|\alpha\alpha' \ket$, $S$ being the $S$-matrix,
 where only initial states  indices appear explicitly (see fig. \ref{f1}). 

At the leading level, all order resummation in the soft-collinear region 
is obtained by a simple expression  that involves
the t-channel total isospin $T$ that couples indices  $\alpha,\beta$:
\be\label{llll}
\ov^H\to \ov^{resummed}= e^{-\frac{\alpha_W}{4\pi}[T(T+1)]
\log^2\frac{s}{M_w^2}}
\;\;\ov^H
\ee
$\alpha_W$ being the weak coupling, $\ov_H$ the hard overlap matrix written in
terms of the tree level $S$-matrix and  $M_w\sim M_z$.

At subleading order, the dressing  by soft and/or collinear radiation
is described at all orders by infrared evolution equations, that are 
$T$-diagonal as far as fermions and transverse gauge bosons 
are concerned \cite{col}. In order to write down the evolution equations
for the case of initial left fermions, we first consider one loop corrections.
At the one loop level, virtual and
real corrections in NLL approximation can be
written as:

\begin{equation}\label{cicciotto}
           \begin{array}{cc}  
\delta\overlap{L}\!_{\alpha\beta}&=
\frac{\alpha_W}{2\pi}\int_{M^2}^s\frac{d\kt^2}{\kt^2}
\int_0^{1}
\frac{dz}{z}\left\{
P^R_{ff}(z)\;\theta(1-z-\frac{\kt}{\sqrt{s}})
\;t^A_{\beta\beta'} \;t^A_{\alpha'\alpha}\; \overlap{L}\!^H\!\!\!_{\alpha'\beta'}\!(zp)\;+
\right.\\
&\left.
P^R_{gf}(z)\;[t^B \,t^A]_{\beta\alpha} \;\overlap{g}\!_{AB}^H(zp)
+ C_f \;
P^V_{ff}(z,\frac{\kt}{\sqrt{s}})\; \overlap{L}\!_{\alpha\beta}^H(p)
\right\}
\end{array}\ee
where
$ P^R_{ff}(z),\,P^R_{gf}(z)$ and $P^V_{ff}(z,\kt) $ 
are defined in \cite{fullend}. The indices   below the overlap matrix label
the kind of particle:
$L=$ Left fermion and $g=$ gauge boson ;
 indices $\alpha,\beta$ refer to the isospin index  ($\alpha=1$ corresponds
 to $\nu$, $\alpha=2$ to $ e$) of the
 lower legs while upper legs
indices are omitted. 

\begin{figure}
      \centering
      \includegraphics[height=80mm]
                  {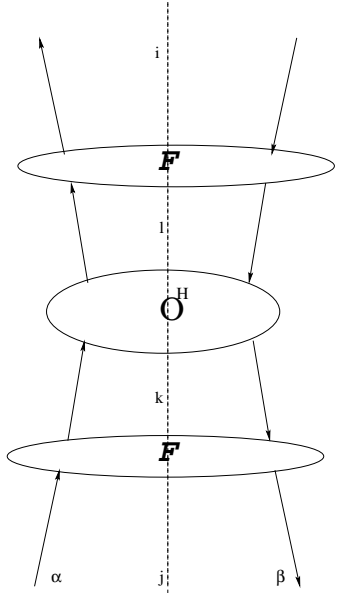}
     
 \caption{\label{f1}
Graphical picture of the factorization formula \ref{fact}.)
 }  
\end{figure}

The one loop formula (\ref{cicciotto}) is consistent
 with a general factorization formula
of type (see fig. \ref{f1}))
\be\label{fact}
\mathop{{\vspace{3pt}\cal O}}\limits^{ i}_{j}
(p_1,p_2;k_1,k_2)=\int \frac{dz_1}{z_1}
\frac{dz_2}{z_2}
\sum_{k,l}^{L,R,g}
\mathop{{\vspace{3pt}\cal F}}\limits^{k}_{j}(z_1;s,M^2)
\mathop{{\vspace{3pt}\cal O}}\limits^{l}_{k}\!^H
(z_1 p_1,z_2 p_2;k_1,k_2) \F{i}{l}(z_2;s,M^2)
\ee
where $i,j$ label the  kind of particle ($L$=left fermion,
$g$=gauge  boson),
 and  where
isospin flavor indices in the overlap function $\ov$ and 
structure function ${\cal F}$ are 
understood.

If the factorization formula (\ref{fact}) is assumed to be valid 
at higher orders as well,
the structure functions will satisfy evolution equations with respect to 
an infrared-collinear  cutoff
$\mu$ parameterizing the lowest value of $\kt$, as follows ($t=\log\mu^2$):
\be\label{kpil}
\der\F{\scriptscriptstyle i}{j}\!_{\alpha\beta}=
\frac{\alpha_W}{2\pi}\left\{
C_f \F{\scriptscriptstyle i}{j}\!_{\alpha\beta} \otimes P^V_{ff}
+[t^C\F{\scriptscriptstyle i}{j}\,^t\; t^C]_{\alpha\beta}\otimes P_{ff}^R
+[t^B\;t^A]_{\beta\alpha}\; \F{\scriptscriptstyle i}{g}\!_{AB}
\otimes P^R_{gf}
\right\}
\ee
In these equations $t^A$ denote the isospin matrices
 in the fundamental
representation and $\F{i}{j}\!{\scriptscriptstyle\alpha\beta}$ 
denotes the distribution of a
particle $i$ (whose isospin indices are omitted)
inside particle $j$ (with isospin
leg indices ${\alpha,\beta}$).
$\F{\scriptscriptstyle i}{j}\,^t$ is the transpose
matrix $\F{i}{j}\!{\scriptscriptstyle\beta\alpha} $. 
Furthermore, we have defined
the convolution
$
[f\otimes P](x)\equiv\int_x^1 P(z)f(\frac{x}{z})\frac{dz}{z}
$.
Since the index $i$ is always kept fixed in (\ref{kpil}), 
it can be omitted,
with the understanding that, for instance, 
$\mathop{{\vspace{3pt}\cal F}}\limits^{}_{j}$ collectively
denotes 
all $\mathop{{\vspace{3pt}\cal F}}\limits^{i}_{j}$ with any 
value of $i$. 

Eqn. (\ref{kpil}) is a matricial   evolution equation; in order to make it
useful the corresponding scalar equations must be written. This can be done by
exploiting the $SU(2)_L$ symmetry which allows to classify the states
according to their isospin quantum numbers. By coupling  the lower legs
$\alpha, \beta$ in fig. 1 one obtains the t-channel isospin eigenstates:
\be
|{\bf T}=0\ket=\frac{1}{\sqrt{2}}(|\nu\nu^*\ket+|ee^*\ket)\qquad
|{\bf T}=1\ket=\frac{1}{\sqrt{2}}(|\nu\nu^*\ket-|ee^*\ket)
\ee
which have $T_L^3=0$ since cross sections always have a given particle on
leg $\alpha$ and its own antiparticle on leg $\beta$. I  now project the
structure operators ${\cal F}$ on these states, omitting the upper leg
indices:
\be\label{xcqef9}\begin{array}{c}
\ff{L}{(0)}=\frac{\bra \nu\nu^*+ee^*|\F{}{L}|\ket}{2}=
\frac{\effe{L}\!_{\nu\nu}+\effe{L}\!_{ee} }{2}  
=\frac{1}{2} {\rm Tr}\left[\F{}{L}\right]
\\
\ff{L}{(1)}=\frac{\bra \nu\nu^*-ee^*|\F{}{L}|\ket}{2}=
\frac{\effe{L}\!_{\nu\nu}-\effe{L}\!_{ee}}{2}=
{\rm Tr}\left[t^3\F{}{L}\right]
\end{array}\ee
Last step in eqs. (\ref{xcqef9}) represents a
 convenient way  to extract the scalar coefficients 
$ \mathop{{\vspace{3pt}f}}\limits^{}_{j}\!^{(T)}$ from 
 $\mathop{{\vspace{3pt}\cal F}}\limits^{}_{j}$; namely,   
by taking  appropriate traces with respect to the
soft leg $j$. For instance $\ff{L}{\scriptstyle(0)}$ 
corresponds to $\frac{1}{2}(
\mathop{{\vspace{3pt}\cal F}}\limits_{L}\!_{ee}+
\mathop{{\vspace{3pt}\cal F}}\limits_{L}\!_{\nu\nu})$  \cite{col} and can
be obtained by 
$\mbox{Tr}_j[\mathop{{\vspace{3pt}\cal F}}\limits^{}_{j}]$; 
here and in the following 
 the trace is taken
 with respect to the indices of the soft lower scale leg
$j$ .
Notice that since gauge and mass eigenstates do not necessarily coincide,
one has to introduce ``mixed legs'' with particles belonging to different
gauge representations on leg $\alpha$ and $\beta$. We label these cases by
$i=LR$ for the mixed left/right fermion leg, 
$i=B3$ for the mixed $W_3-B$ gauge bosons and
$i=h3$ for the Higgs sector case. These mixing phenomena are interesting by
themselves and have been considered in \cite{full,lon,abelian,transverse} at double log level.

Projecting eq. (\ref{kpil}) for instance 
on the $T=0$ component we obtain:
\be
\der\mbox{Tr}[\F{}{f}]=
\frac{\alpha_W}{2\pi}\left\{
C_f \mbox{Tr}[\F{}{f}] \otimes P^V_{ff}
+\mbox{Tr}[t^C\F{}{f}\,^t\;t^C]\otimes P_{ff}^R
+\mbox{Tr}[t^B\,t^A]\;\F{}{g}\!\!_{AB}\;\otimes P^R_{gf}
\right\}
\ee
where the traces are taken, here and in the following,
with respect to the soft leg indices. This 
gives:
\be\label{andra}
\der \ff{L}{(0)}=
\frac{\alpha_W}{2 \pi}\;\left(
\frac{3}{4}\; \ff{L}{(0)}\otimes \left( P^V_{ff}+P^R_{ff} \right)+
 \frac{3}{4}\, \ff{g}{(0)}\otimes P^R_{gf}\right)
\ee
after taking into account that 
$\mbox{Tr}[t^B\;t^A]\,\F{}{g}\!\!_{AB}=\frac{1}{2}\sum_A\F{}{g}\!_{AA}=
\frac{1}{2}\mbox{Tr}[\F{}{g}]$ and
that $\ff{g}{(0)}=\frac{1}{3}\mbox{Tr}[\F{}{g}]$.

Finally, the last step in obtaining the all order resummed overlap matrix,
requires the evolution of the
$\ff{i}{(T)}$'s according to eqn. (\ref{andra}) with appropriate initial
conditions, and inserting the evolved 
$\ff{i}{(T)}$'s into (\ref{fact}). This can by done by exploiting the 
recovered isospin symmetry, which allows us to write: 
\be
\mathop{{\vspace{3pt}\cal O}}\limits^{ i}_{j}
(p_1,p_2;k_1,k_2)=\sum_{T}
\int \frac{dz_1}{z_1}
\frac{dz_2}{z_2}
\sum_{k,l}^{L,R,g}\;
\mathop{{\vspace{3pt} f}}\limits^{k}_{j}\!_{\!\!(T)}(z_1;s,M^2)
\;\mathop{{\vspace{3pt}\cal O}}\limits^{l}_{k}\!^H_{(T)}
(z_1 p_1,z_2 p_2;k_1,k_2)\;
\mathop{{\vspace{3pt} f}}\limits^{i}_{l}\!_{\!\!(T)}(z_2;s,M^2)
\ee
\section{Electroweak evolution equation in the full Standard Model}
The set of complete infrared evolution equation in the electroweak sector
of the Standard Model have been derived in \cite{fullend}.
According to the equivalence theorem, we replace longitudinal
gauge bosons with the corresponding Goldstone bosons. 
We choose to work in an axial gauge so that this substitution can be done
without higher order corrections  in the definition of the asymptotic states
\cite{Beenakker:2001kf}.
The procedure of writing down evolution equations for the matricial
$\F{i}{j}\!_{\alpha'\beta'}^{\alpha\beta}$
and then extracting the corresponding equations for the scalar components 
$\mathop{{\vspace{3pt}f}}\limits_{i}^{j}$ is similar to the one outlined in
previous section. However a number of contributions has to be added: the
ones proportional to $g'$, and the ones proportional to the third
generation Yukawa couplings, which cannot be neglected even in the high
energy limit. While the various contributions are summarized in fig, I
refer to \cite{fullend} for the complete set of equations.
\begin{figure}
      \centering
      \includegraphics[height=90mm]
                  {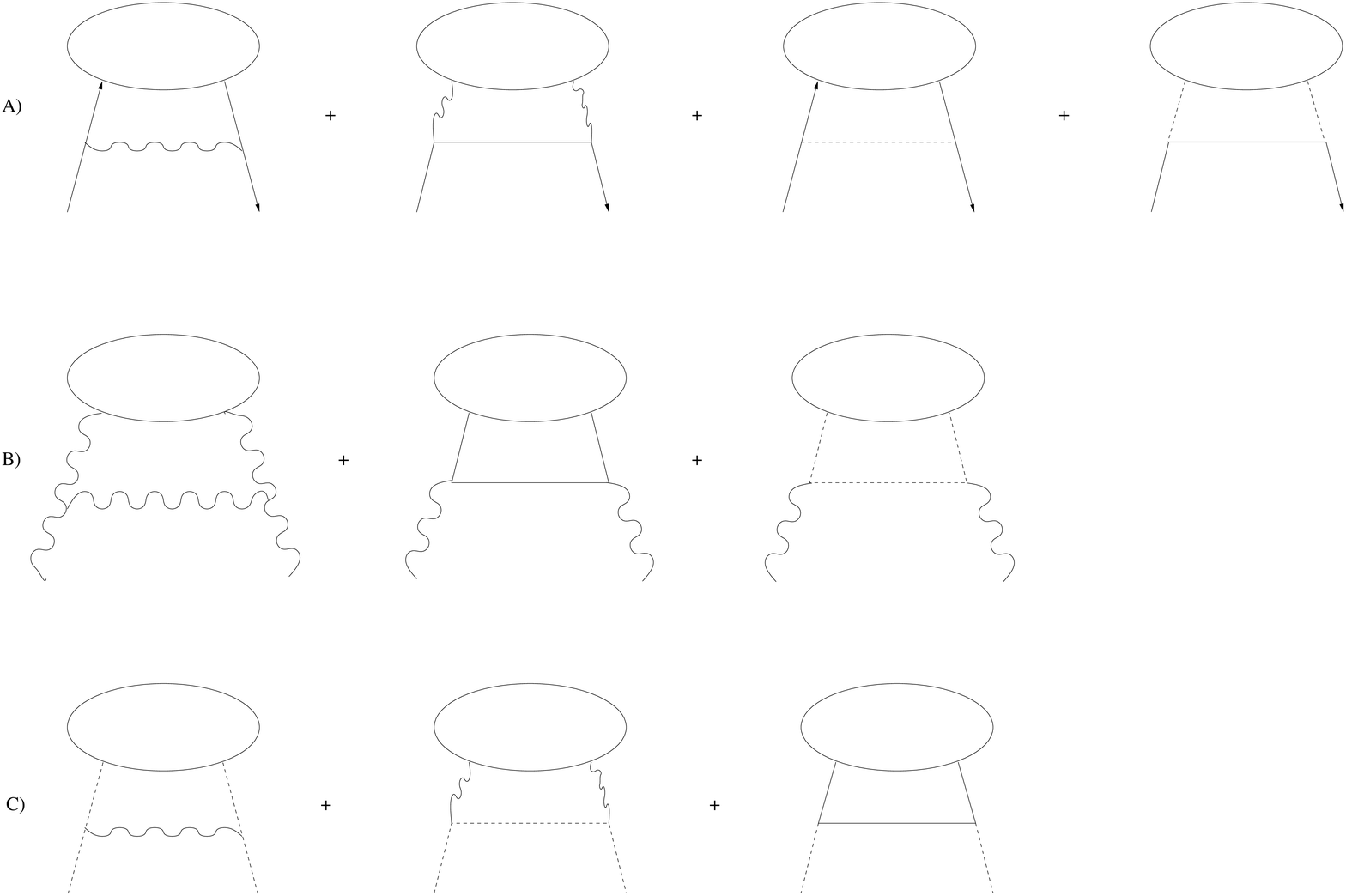}
     
 \caption{\label{figg}
Leading real emission Feynman diagrams in axial gauge:
A) Feynman diagrams contributing to the evolution of the fermionic
structure functions;
B) Feynman diagrams contributing to the evolution of the  transverse
gauge boson structure functions;
C) Feynman diagrams contributing to the evolution of the scalar 
structure functions.
The wavy lines are  transverse gauge bosons, 
dashed lines stay for Higgs sector particles and 
straight lines for fermions.
 }  
      \end{figure}
A further complication is due to the fact that, in order to 
provide a complete classification
of the states, the conserved quantum number $CP$ is needed on top of 
the SU(2)$\otimes$ U(1) quantum numbers.

Of course, the task of writing evolution equations for the full Standard
Model is still incomplete. Having in mind hadronic initial states like in
the LHC for instance, one has to add QED effects at low ($\mu< M$) scale,
and QCD effects. Moreover, a careful treatment of matching conditions at
the weak scale $M$ is mandatory. These will be the key points of our future
research line.

\end{document}